\documentclass[twoside]{article}
\usepackage{fleqn,espcrc2}
\usepackage{graphicx}
\usepackage{here}%pour inserer figures et tables a emplacement

\newcommand{\AmS}{{\protect\the\textfont2
    A\kern-.1667em\lower.5ex\hbox{M}\kern-.125emS}}
\newcommand{\be}{\begin{equation}}
\newcommand{\ee}{\end{equation}}
\newcommand{\bea}{\begin{eqnarray}}
\newcommand{\eea}{\end{eqnarray}}

\def\beq{\begin{equation}}
\def\eeq{\end{equation}}
\def\bea{\begin{eqnarray}}
\def\eea{\end{eqnarray}}
\def\bq{\begin{quote}}
\def\eq{\end{quote}}

\def\nnb{\nonumber}
\def\ga{\left(}
\def\dr{\right)}

\def\rar{\rightarrow}
\def\Lrar{\Longrightarrow}

\def\nnb{\nonumber}
\def\la{\langle}
\def\ra{\rangle}
\def\nin{\noindent}
\def\ba{\begin{array}}
\def\ea{\end{array}}

\def\bl{\bullet}

\def\als{\alpha_s}

\def\g2{ \la\alpha_s G^2 \ra}
\def\g3{g^3f_{abc}\la G^aG^bG^c \ra}
\def\g4{\la\als^2G^4\ra}

\title{\bf{
$V$-$A$ hadronic
tau decays : a laboratory for the QCD vacuum }} 
\author{
Stephan Narison\address{Laboratoire de Physique Th\'eorique et d'Astrophysique,
CNRS \& Universit\'e de Montpellier 2,
Place Eug\`ene Bataillon,
34095 - Montpellier Cedex 05, France \\ 
E-mail:
snarison@yahoo.fr}
}
\begin{document}
\begin{abstract}
\nin
ALEPH/OPAL data on the $V$-$A$ spectral functions from hadronic $\tau$ decays are
used in connection with a set of Laplace transform sum rules (LSR) for fixing the size of the QCD vacuum
condensates up to dimension 18. Our results favour the ones from large-$N_c$ QCD within the
Minimal Hadronic Approximation (MHA) and show a violation of about a factor 2-5
of the vacuum saturation estimate of the dimension-six to -ten condensates. We scrutinize the different determinations of the 
QCD vacuum condensates using $\tau$-decays data. After revisiting some of the existing results, we present coherent values of 
the condensates from different methods.
%\vspace{-4.2cm}
\end{abstract}
\maketitle
\pagestyle{plain}
%\maketitle
\section{INTRODUCTION}
\nin
Hadronic tau decays have been demonstrated \cite{BNP} to be an
efficient laboratory for testing perturbative and non-perturbative QCD. That is due both to the
exceptional value of the tau mass situated at a frontier regime between
perturbative and non-perturbative QCD and to the excellent quality of the
ALEPH/OPAL \cite{ALEPH,OPAL} data. On the other, it is also known before the
advent of QCD, that the Weinberg \cite{WEINBERG} and DMO \cite{DMO} sum rules
are important tools for controlling the chiral and flavour symmetry realizations
of QCD, which are broken by light quark mass terms to higher order \cite{FNR}
and by higher dimensions QCD condensates \cite{SNWEIN} within the SVZ expansion
\cite{SVZ}\footnote{For a review, see e.g. \cite{SNB}.}. \\ For completing our program in the vector and $V+A$ 
channel~\cite{ZAKA1,ZAKA2,SNG,PICH2,TARRACH}, we
probe, in this paper,  the structure of the QCD vacuum using the ALEPH/OPAL data on the
$V$-$A$ spectral functions in connection with a set of Laplace sum rules (LSR).
We have already initiated the analysis of the $V-A$ channel in previous papers \cite{SNL1,SNL2}. However, our
main motivation here is due to the recent interests on the
$V-A$ hadronic correlator, which can serve as an order parameter of spontaneous chiral
symmetry breaking in the chiral limit $m_q=0$. This correlator also governs the dynamics of the
weak matrix elements of the electroweak Penguin-like operators \cite{ENJL,DONO}. These
important properties require a detailed structure of the related QCD vacuum which can be
parametrized by the sum of power corrections \cite{SVZ}. A large number of papers on the
estimates of these power corrections using  different methods exist in the literature,
but with conflicting results in \cite{ALEPH,OPAL,RAF04,PERIS,STERN,ZYA,IZ}
and in
\cite{PRADES,DOMINGUEZ,ROJO,MALT,CIULLI}. In the following we propose a set of Laplace transform sum
rules (LSR) which can help to clarify such discrepancies. 
We shall be concerned here with the $V$-$A$ two-point correlator:
\bea
&\Pi_{\mu\nu}^{LR}(q)\equiv i\int d^4 x~ e^{iqx}\la 0|{\cal T} J_\mu^L(x)\ga
J_\nu^R(0)\dr^{\dagger}|0\ra\nnb\\
&\stackrel{m_q\rar 0}{=}-(g_{\mu\nu}q^2-q_\mu q_\nu)\Pi^{LR}(q^2)~,
\eea
built from the left-- and right--handed components of the local weak current:
\beq
J_\mu^{L}=\bar u\gamma_\mu(1-\gamma_5)d,~~~~~~~J_\mu^{R}=\bar
u\gamma^\mu(1+\gamma_5)d~.
\eeq
Following SVZ \cite{SVZ}, the correlator can be approximated by:
\beq
\Pi^{LR}(Q^2)\simeq\sum_{d\geq 2}{{\cal O}_{2d}\over (Q^2)^d}
\eeq
where ${\cal O}_{2d}\equiv C_{2d}\la{\cal O}_{2d}\ra$ is the short hand-notation of the the QCD non-perturbative
condensates $\la{\cal O}_{2d}\ra$ of dimension
$D\equiv 2d$ and its associated perturbative Wilson coefficient $C_{2d}$; $q^2\equiv -(Q^2 >0)$ is the momentum transfer.
In the chiral limit $m_{u,d}=0$, there is no $D=2$ term as unflavored contribution of the renormalon-type
\cite{ZAKA,ZAKA1,ZAKA2} vanishes. The spectral function $(v-a)$:
\beq
\frac{1}{\pi}{\rm Im}\Pi^{LR}\equiv\frac{1}{2\pi^2}\ga v-a\dr~.
\eeq
has been measured by
ALEPH and OPAL \cite{ALEPH,OPAL} using
$\tau$-decay data.  Within a such normalization, the original ``sacrosante" first and second
Weinberg sum rules \cite{WEINBERG,DMO} read, in the chiral limit $m_{u,d}=0$:
\bea
{\cal S}_{0}&\equiv&\int_0^{\infty} dt ~\frac{1}{\pi}{\rm Im}\Pi_{LR} -
2f^2_\pi=0~,\nnb\\
{\cal S}_{1}&\equiv&\int_0^{\infty} dt ~t~\frac{1}{\pi}{\rm Im}\Pi_{LR} =
0~,
\eea
where $f_\pi=(92.4\pm 0.26)$ MeV is the experimental pion decay
constant. 
%%%%%%%%%%%%%%%%%%%%%%%%%%%%%%%%%%%%%
\section{THE LAPLACE SUM RULES (LSR)}
\nin
In order to exploit the ALEPH/OPAL
\cite{ALEPH,OPAL} data on the spectral function $v-a$ from hadronic tau--decays, we shall work
with the LSR version of the 1st Weinberg sum rule, in the chiral limit $m_{u,d}=0$:
\bea
{\cal L}_0(\tau)&=&\int_0^{\infty} dt ~{\rm e}^{-t\tau}~\frac{1}{\pi}{\rm Im}\Pi_{LR}-2f_\pi^2\nnb\\
&\simeq&
\sum_{d\geq 2}{\tau^{(d-1)}\over (d-1)!}~{\cal O}_{2d}~, 
\eea
from which on can obtain, by taking successive derivatives in $\tau$, the set of LSR:
\bea
{\cal L}_n&\equiv&(-1)^n{d^n{\cal L}_0\over d\tau^n}\simeq \int_0^{\infty} dt ~t^n~{\rm e}^{-t\tau}~\frac{1}{\pi}{\rm
Im}\Pi_{LR}\nnb\\ &\simeq&
(-1)^n\sum_{d\geq (n+1)}{\tau^{(d-n-1)}\over (d-n-1)!}~{\cal
O}_{2d}~,
\eea
For our purpose, we shall truncate (in order to have a much better comparison with the existing results) the series at $2d=18$-dimension
condensates \footnote{The result will not depend crucially on the choice of the truncation of the series. We shall see that at the region where the
condensates are estimated the OPE presents a good convergence.}, assuming that this approximation provides a good description of the exact expression of the two-point correlator $\Pi^{LR}$.
Then, from our previous general formula, one can write the set of sum rules:
\bea\label{eq: clsr}
{\cal L}_8&\simeq&
+{\cal
O}_{18}\nnb\\
{\cal L}_7&\simeq&
-{\cal
O}_{16}-{\cal
O}_{18}\tau\nnb\\
{\cal L}_6&\simeq&
+{\cal
O}_{14}+{\cal
O}_{16}\tau+{\cal
O}_{18}{\tau^2\over 2}\nnb\\
.~.~.&&~~~.~.~.~~~~.~.~.~~~~~~.~.~.
\eea
%%%%%%%%%%%%%%%%%%%%%%%%%%%%%
%%%%%%%%%%%%%%%%%%%%%%%%%%%%%%%%%%%
%\vfill\eject
\begin{table*}[h]
\setlength{\tabcolsep}{.1pc}
\newlength{\digitwidth} \settowidth{\digitwidth}{\rm 0}
\catcode`?=\active \def?{\kern\digitwidth}
%\begin{center}
\caption{Estimated values of the $D\equiv 2d\leq 18$-dimension $\la {\cal O}_{D}\ra$ condensates in units of $10^{-3}$
GeV$^{D}$. We have ordered the condensates from ${\cal O}_{18}$ to ${\cal O}_{4}$ according to their chronological estimate.}
\label{tab: comparison}
\begin{tabular*}{\textwidth}{@{}l@{\extracolsep{\fill}}lcccccccc}
%\begin{tabular}[H]{lcccccccc}
\hline
\hline
& & &&&&\\
Authors&${\cal O}_{18}$&${\cal O}_{16}$&${\cal O}_{14}$&${\cal O}_{12}$&${\cal O}_{10}$&${\cal O}_{8}$&${\cal
O}_{6}$&${\cal O}_{4}$\\ &&&&\\
\hline
\hline
&&&&\\
{\bf THIS WORK}&&&&\\
\\
{\bf LSR} \\ 
Eq. (\ref{eq: clsr})&$-1\pm 0.6$&$+4.3\pm 1.9$&$-9.6\pm 3.1$&$+14.7\pm 3.7$&$-17.1\pm 4.4$&$+15.4\pm 4.8$&$-9.7\pm
4.1$&$-0.5\pm 0.1$\\
Eqs. (\ref{eq: ilsr1},~\ref{eq: ilsr2})&&&&&&$+15.8\pm 3.2$&$-8.4\pm 1.6$&
\\
Eq. (\ref{eq: taulike})&&&&&&&$-8.0\pm 1.1$&
\\ 
 {\bf Average} &&&&&&\boldmath$ +15.6\pm  4.0$ &\boldmath$ -8.7\pm  2.3$&\\
\\
{\bf wFESR}\\
{Rev.}\footnote{...}&&$+30\pm 10$&$-28\pm 8$&$+25\pm 5$&$-22\pm 3$&$+16.8\pm 2.0$&$-10.2\pm .4$\\
\it\footnotesize Orig.\cite{MALT}\footnote{...}&&\footnotesize -$\it 946\pm 147$&\footnotesize +$ \it 390\pm 65$&
\footnotesize -$\it 146\pm 27$&\footnotesize +$ \it 43.5\pm 10.5$&\footnotesize -$\it 4.4\pm
3.8$&\footnotesize -$\it 4.8\pm 0.9$\\
%&&&\\
\\
{\bf FESR}\footnote{...}\\
BG Rev.\footnote{...} &&&&&&$+12\pm 1.5$&$-6.6\pm 0.2$\\
\it\footnotesize BG Orig.\cite{PRADES}&&&&&&\footnotesize $\it -12.4\pm 9.0$&\footnotesize $\it -3.2\pm
2.0$\\
\\
DS Rev.\footnote{...}&&&&&&$ +10\pm 2$ &$-8.0\pm 2.0$\\
\it\footnotesize DS Orig.\cite{DOMINGUEZ}&&&&&&\footnotesize$\it -2\pm 12$ &\footnotesize$\it -8.0\pm
2.0$\\
\\
{LR} Rev.\footnote{...}&&&&$+53\pm 16$&$-39\pm 12$&$ +26.0\pm 8.4 $&$ -14.7\pm 4.8$\\
\it\footnotesize LR Orig.\cite{ROJO}&&&&\footnotesize$\it -260\pm 80$&\footnotesize$\it +78\pm
24$&\footnotesize$
\it -120^{+7}_{-11}
$&
\footnotesize$\it -4\pm 2$\\ &&&\\
\hline
&\\
{\bf OTHERS}\\
\\
MHA+$\rho'$\cite{RAF04}& &$+11.5\pm 3.5$&$-12.5\pm 3.4$&$+13.2\pm 3.3$&$-13.1\pm 3.0$&$+11.7\pm 2.6$&$-7.9\pm 1.6$\\ 
MHA\cite{RAF04}& &$+11.9\pm 3.9$&$-12.8\pm 3.9$&$+13.3\pm 3.9$&$-13.2\pm 3.6$&$+11.7\pm 3.1$&$-7.9\pm 2.0$\\ 
ZYA\cite{ZYA}&&&&&$-4.5\pm 3.4$&$+7.8\pm 3.0$&$-7.1\pm 1.5$\\
IZ\cite{IZ}&&&&&&$+7.0\pm 4.0$&$-6.4\pm 1.6$\\
ALEPH\cite{ALEPH}&&&&&&$+11.0\pm 1.0$&$-7.7\pm 0.8$\\
OPAL\cite{OPAL}&&&&&&$+7.5\pm 1.3$&$-6.0\pm 0.6$\\
DGHS\cite{STERN}&&&&&&$+8.7\pm 2.4$&$-6.0\pm 0.6$\\
CS3\cite{CIULLI}&&&&&&&$-4.0\pm 2.8$\\
\\
\hline\\
{\bf AVERAGE} \footnote{...} &$-1\pm 0.6$&$+14.4\pm 4.6$&$-15.7\pm 3.7$&$+23.8\pm 6.4$&$-18.2\pm
5.9$&$12.2\pm 2.9$&$-7.8\pm 1.6$\\
\\
\hline
\hline
\end{tabular*}
{\footnotesize 
\begin{quote}
$^{3}\,$ We have redone the analysis of \cite{MALT} using $t_c$-stability criterion. \\
$^{4}\,$ We use the mean value of the results from the ALEPH and OPAL data.\\
$^{5}\,$ The revised (Rev.) FESR results have been obtained at
$t_c\approx 1.5$ GeV$^2$; the original (Orig.) ones at $t_c\approx$ 2.5 GeV$^2$.\\
$^{6}\,$ These results have been obtained by \cite{PRADES} at 1.5 GeV$^2$.\\
$^{7}\,$ We have corrected the value of ${\cal O}_8$ (see section \ref{sec: fesr}) and rescaled the results of
\cite{DOMINGUEZ}.\\
$^{8}\,$ The central values come from \cite{ROJO2}. Inspired from the results of \cite{ROJO} at $2.5$
GeV$^2$, we have roughly estimated the \, $~~~~$systematic errors to be about 30\%. \\
$^{9}\,$ Numbers in the lines {\it Orig.} are not considered into the average.
\noindent
\end{quote}}
%\end{center}
\end{table*}
\nin
%%%%%%%%%%%%%%%%%%%%%%%%%%%%%%%%%%%%%

%\vspace{-1.5cm}
\begin{figure*}[hbt]
\begin{center}
\includegraphics[width=14cm]{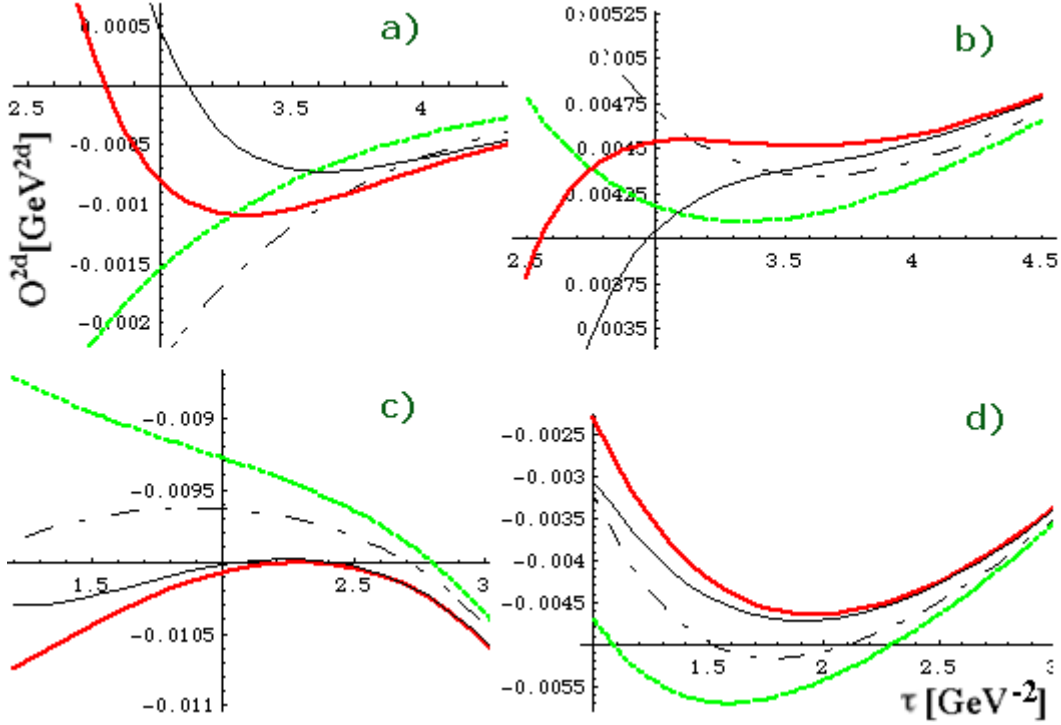}  
\vspace{-0.5cm}
\caption{ \footnotesize $\tau$ in GeV$^{-2}$-behaviour of the condensates ${\cal O}_{2d}$ in units of GeV$^{2d}$ for different values of $t_c$
in GeV$^2$: 1.4 (dot-dashed), 1.5 (dashed bold), 2.5 (continuous bold), 2.6 (continuous): a) ${\cal O}_{18}$, b) ${\cal O}_{16}$, c) ${\cal
O}_{6}$ and d)
${\cal O}_{4}$.}
\label{fig: d64}
\end{center}
\end{figure*}
%\vspace{-1cm}
\nin
%%%%%%%%%%%%%%%%%%%%%%%%%%%%%%%%%%%%%%%%%%
$\bl$ Therefore, we can extract iteratively the vacuum condensates beginning from ${\cal L}_8$. The
value of ${\cal
O}_{18}$ obtained in this way will be inserted into ${\cal L}_7$ for determining 
${\cal
O}_{16}$ and so on.\\
$\bl$ We parametrize the spectral function by using the ALEPH/OPAL
\cite{ALEPH,OPAL} data on the spectral function $v-a$ from hadronic tau--decays below $t_c$. Above $t_c$, we use a QCD continuum 
coming from the discontinuity of the QCD diagrams. In the particular case of ${\rm Im}\Pi^{LR}$ studied here, a such contribution vanishes
identically, which is equivalent to cut the integral in Eq. (\ref{eq: clsr}) at $t_c$. The appropriate values of
$t_c$ has been studied in
\cite{PERIS,SNL1,PRADES,RAFAEL} by requiring that the 1st and the 2nd Weinberg sum rules vanishes in the chiral limit to leading order of the
OPE. Two solutions have been found:
\beq\label{weinberg}
t_c\simeq (1.4\sim 1.5)~{\rm GeV}^2~~~{\rm and}~~~
(2.5\sim 2.6)~{\rm GeV}^2.
\eeq
In \cite{SNL1} the lowest value of $t_c$ has been favoured due to the
inaccuracy of the ALEPH/OPAL data which affects the highest value, though intuitively, one tends to favour
this highest value of $t_c$ where pQCD is expected to work better. Exluding the high-$t_c$ value solution, and requiring simultaneous
zeros of the 1st and 2nd Weinberg sum rules, Ref. \cite{SNL1} deduces the accurate number:
\beq\label{tcoptimal} 
t_c=(1.475\pm 0.015)~{\rm GeV}^2.
\eeq
This choice, as emphasized in \cite{PERIS} co\"\i ncides with the $t_c$-value obtained for the MHA in the
large $N_c$-limit, which follows from the duality relation \cite{RAFAEL}:
\beq
t_c\simeq 8\pi^2f_\pi^2{1\over 1-g_A}\simeq (1.2\pm 0.2)~{\rm GeV}^2~,
\eeq
with $g_A\simeq 0.5\pm 0.06$. Instead in \cite{PRADES,DOMINGUEZ,ROJO}, the
higher value of
$t_c$ around 2.5 GeV$^2$ has been favoured.\\
$\bl$ In order to avoid results which strongly depend on these choices of $t_c$, we only consider the
above values of $t_c$ as a guideline of our analysis. Indeed, it is unlikely to take
$t_c\leq 1.4$ GeV$^2$ as we will loose part of the $\rho$ meson tails, and then most of the lowest ground state
dynamics. Taking $t_c\geq 2.6$ GeV$^2$, the kinematic region is small and the data become very inaccurate.
Then, they cannot provide useful information to the spectral function. Indeed in this region, the spectral
function does not have a definite sign, for a given data point, due to the large error bars.\\
$\bl$ For an illustration, we show the analysis of ${\cal O}_{18,16}$ and ${\cal O}_{6,4}$  in 
Fig. \ref{fig: d64}  for different choices of the $t_c$-cut until which we use the ALEPH/OPAL
data, and beyond which the pQCD diagram is expected to describe the two-point
correlator. The analysis of the other condensates present similar features.\\
$\bl$  The optimal results  given in Table
\ref{tab: comparison} correspond to the one at the minimum or inflexion point of
$\tau$ for different $t_c$-values inside the range in Eq. (\ref{weinberg}). The $\tau$-stability criterion
has been often used in the Laplace sum rules analysis as it signals the {\it compromise region} where the OPE is reliable (smaller
$\tau$-values) and where the information from the data still remains optimal (larger $\tau$-values). It is also unlikely if the result is
strongly dependent on the choice of $t_c$-values as this signals a strong model dependence of the result on the form of the QCD continuum.
Then, in the following, we shall use in connection these two stabilities criteria for extracting the optimal results\footnote{For more
complete discussions, see e.g. see e.g.
\cite{SNB}.}.
 \\
$\bl$ The error takes into account the one of the
data and the systematics of the method due to the range of $t_c$-values given in Eq.
(\ref{weinberg}) and to the propagation of errors induced by the ones of the input condensates. We do not include some eventual
statistical errors.\\
$\bl$ It is important to notice from our analysis that in the range of $t_c$ given in Eq. (\ref{weinberg}), the extracted values
of the condensates do not flip sign contrary to the case of FESR's results given in the existing literature. One can attribute
this feature to the role of the exponential weight in LSR which enhances the contribution of the low-energy region to the sum
rule.\\
$\bl$ One
can also notice that, for high-dimension condensates, the optimal values are obtained at large $\tau$-values like also in the
least square fit analysis of \cite{ZYA,IZ}. However, we have checked that, during the analysis of each sum rule, the
high-dimension condensates remain corrections to the low-dimension contributions and do not break the OPE. One
can also notice that the position of the minimum shifts to lower values of $\tau$ for
decreasing dimension condensates, as one can see in Fig. \ref{fig: d64} for the $D=18$
to the $D=4$ condensates. These features are  re-assuring for the reliability of the result.\\
$\bl$ In order to test the accuracy of our estimate, we have extracted from ${\cal L}_0$ the known tiny value of the ${\cal O}_4$
quark condensate contribution using as input all higher dimension condensates. Including radiative corrections, this contribution
reads
\cite{SVZ,GENERALIS}:
\beq
{\cal O}_4^{th}=2(m_u+m_d)\la\bar uu\ra \Bigg{[} 1+{4\over 3}{\alpha_s\over\pi}+{59\over
6}\ga{\alpha_s\over\pi}\dr^2\Bigg{]},
\eeq
where $(m_u+m_d)\la\bar uu+\bar dd\ra=-2f_\pi^2m_\pi^2$. The size of the radiative corrections is about 35\%
at the
$\tau$-scale where the optimal results are extracted. This gives in units 
of $10^{-3}~{\rm GeV}^4$:
\beq\label{eq: gmor}
{\cal O}_4^{th}\simeq -0.44~,
\eeq
\vspace{-0.15cm}
in excellent agreement with our fit $-0.5\pm 0.1$ given in Table \ref{tab: comparison}
from Fig. \ref{fig: d64}d). This test increases the confidence on our other predictions in 
Table \ref{tab: comparison} obtained in the same way.
%%%%%%%%%%%%%%%%%%%%%%%%%%%%%%%%%%%%%%%%%%%%%%%%%%%%%%%%%%%%%%%%%%%%%%%%%%%%%%%%%%%%%%
\section{ALTERNATIVE ESTIMATES OF {\boldmath${\cal O}_6$} AND  {\boldmath${\cal O}_8$}
}\label{sec: improve}
\begin{figure}[hbt]
\begin{center}
\includegraphics[width=7cm]{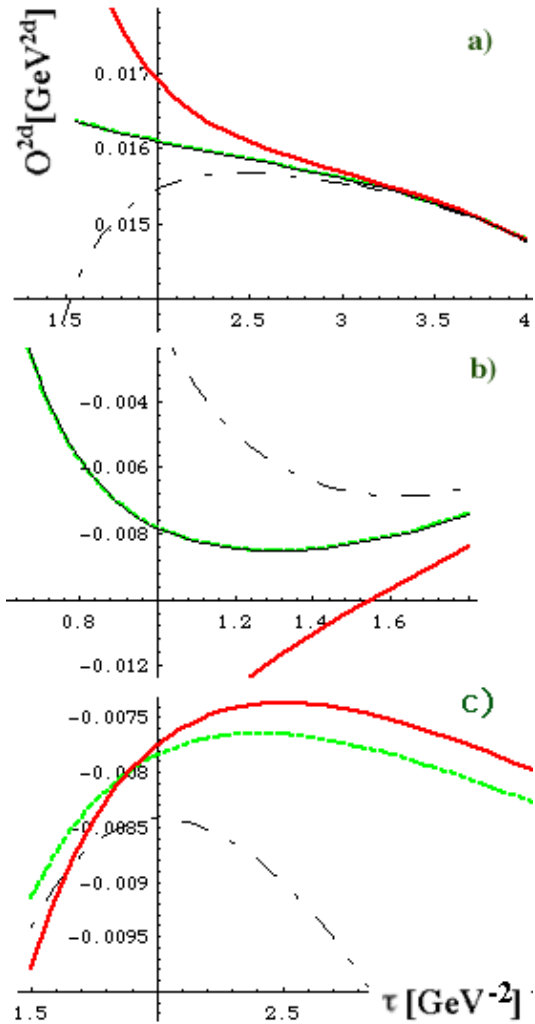}  
\vspace{-0.5cm}
\caption{ \footnotesize Same as in Fig. \ref{fig: d64} but for the improved  analysis in section \ref{sec: improve}: a) 
${\cal O}_8$ from Eq. (\ref{eq: ilsr2}), b) ${\cal O}_6$ from Eq. (\ref{eq: ilsr1}) and c) ${\cal
O}_6$ from Eq. (\ref{eq: taulike}).}
\label{fig: d68}
\end{center}
\end{figure}

\nin
Using the previous method, we have obtained from Eq. (\ref{eq: clsr}) the results in Table \ref{tab: comparison} in units of
$10^{-3}~{\rm GeV}^D$ (D being the dimension of the condensates).
Here, we present alternative estimates based on 
some combinations of LSR in the chiral limit $m_{u,d}=0$.\\
$\bl$ The first sum rule is chosen in such a way that ${\cal O}_8$ disappears to leading order while higher
dimensions
$D=10,12$ have smaller coefficients than in the individual sum rules. 
\vspace*{-0.1cm}
\bea\label{eq: ilsr1}
3{\cal L}_0+\tau{\cal L}_1&=&2{\cal O}_4\tau+{\cal O}_6\frac{\tau^2}{2}-{\cal
O}_{10}\frac{\tau^4}{24}-{\cal O}_{12}\frac{\tau^5}{60}\nnb\\
&-&{\cal O}_{14}\frac{\tau^6}{240}-{\cal O}_{16}\frac{\tau^7}{1260}
-{\cal O}_{18}\frac{\tau^8}{8064}.
\eea
$\bl$ In the second sum rule, ${\cal O}_6$ disappears and then ${\cal O}_8$ will dominate the LSR:
\bea\label{eq: ilsr2}
{\cal L}_0+{\tau\over 2}{\cal L}_1&=&{\cal O}_4{\tau\over 2}-{\cal O}_8\frac{\tau^3}{12}-{\cal
O}_{10}\frac{\tau^4}{24}-{\cal O}_{12}\frac{\tau^5}{80}\nnb\\
&-&{\cal O}_{14}\frac{\tau^6}{360}-{\cal O}_{16}\frac{\tau^7}{2016}-{\cal O}_{18}\frac{\tau^8}{13440}~.
\eea
$\bl$ Therefore, we use the sum rule in Eq. (\ref{eq: ilsr1}) (resp. Eq. (\ref{eq: ilsr2})) for extracting
${\cal O}_6$ (resp. ${\cal O}_8$). We use the known
tiny value of $D=4$ quark condensate contribution given in Eq. (\ref{eq: gmor}).
The analysis is shown in Fig.~\ref{fig: d68} and the results are given in Table \ref{tab: comparison}.\\
$\bl$ Finally, we analyze the $\tau$-like decay sum rule, which has the advantage to be kinematically suppressed near
the real axis:
\bea\label{eq: taulike}
{\cal L}_{01}&\equiv&\int_0^{t_c} dt ~\ga 1-{t\over t_c}\dr{\rm e}^{-t\tau}~\frac{1}{\pi}{\rm Im}\Pi_{LR}\nnb\\
&=&\sum_{n\geq 2} {\cal O}_{2n} {\tau^{(n-1)}\over (n-1)!}\Big{[}1+{(n-1)\over t_c\tau}\Big{]}~,
\eea
from which we deduce ${\cal O}_6$ using as input ${\cal O}_4$ and the higher dimension condensates.\\
$\bl$ Our different results are summarize in Table \ref{tab: comparison}. 
%%%%%%%%%%%%%%%%%%%%%%%%%%%%%
%\vspace{-1cm}
\nin
%%%%%%%%%%%%%%%%%%%%%%%%%%%%%%%%%%%%%%%%%%

%%%%%%%%%%%%%%%%%%%%%%%%%%%%%%%%%%
\section{COMPARISON WITH EXISTING ESTIMATES}\label{sec: comparison}
\nin
%%%%%%%%%%%%%%%%%%%%%%%%%%%%%%%%%%%%%%%%%%%%%%%%%%%%%%%%%%%%%%%
\section*{a) Large-$N_c$ and Minimal Hadronic Approximation (MHA)}
\nin
$\bl$ Our results agree in signs and in magnitude until the $D=14$-dimension condensates with the ones in \cite{RAF04}
obtained using large $N_c$ and the minimal hadronic approximation (MHA) and with its improved version including the next radial
vector meson~$\rho'$. \\
$\bl$ Our result for the $D=16$ condensate still agrees in sign with the one in \cite{RAF04} but
our absolute value is lower than the  one in \cite{RAF04} by about $2\sigma$.
%%%%%%%%%%%%%%%%%%%%%%%%%%%%%%%%%%%%%%%%%%%%%%%%%%%%%%%%
\section*{b) ALEPH and OPAL estimates from $\tau$-decay}
\nin
Our results for the low $D=6,8$ condensates agree also quite well with the ALEPH/OPAL estimate of the separate $V$ and $A$
channels \cite{ALEPH,OPAL}, \cite{STERN}, from which one can deduce the $V$-$A$ difference.
%%%%%%%%%%%%%%%%%%%%%%%%%%%%%%%%%%%%%%%%%
\section*{c) Exponential-like sum rules}
\nin
$\bl$ The value of the $D=6$
condensate also agrees within the errors with the results in \cite{ZYA}, but the values of ${\cal O}_{8,10}$ obtained
in the present paper are  about two times higher. However, our analysis differs from \cite{ZYA} who use a least
square fitting procedure with some other forms of LSR with a different kernel. Due to the alternate sign of the condensate
contributions to the OPE, the fitting procedure can be inaccurate as we shall see explicitly in a forthcoming example.

%%%%%%%%%%%%%%%%%%%%%%%%%%%%%
\begin{figure*}
\begin{center}
\includegraphics[width=13cm]{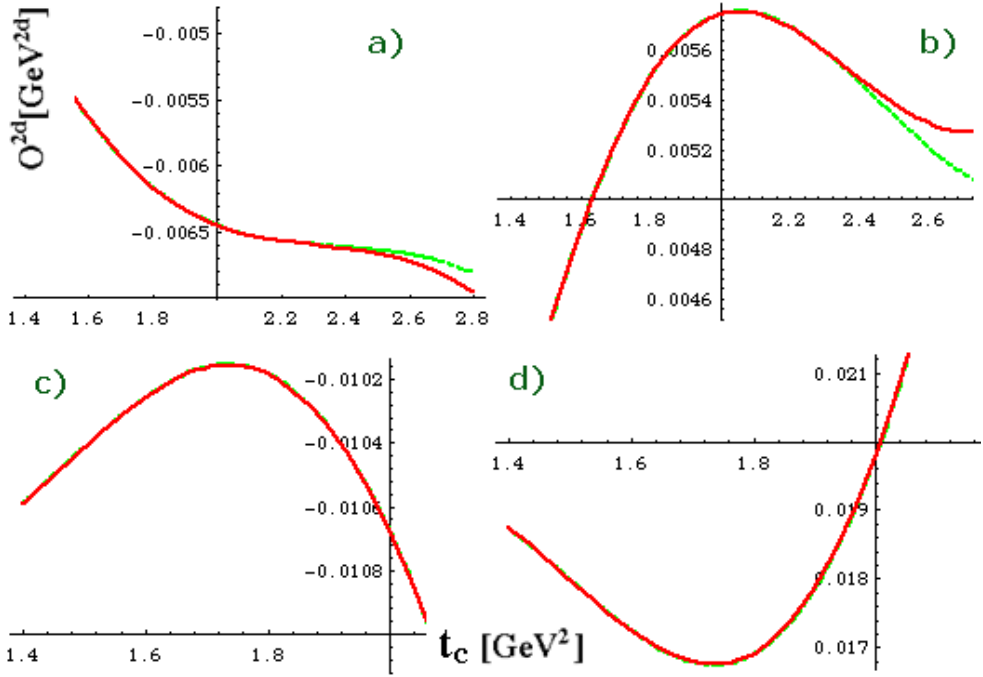}  
\vspace{-0.5cm}
\caption{ \footnotesize $t_c$-behaviour in GeV$^2$ of different observables used in \cite{MALT}: a) $(t_c^2/7)J_{w_1}$, b)
$(t_c^2/2)J_{w_2}$, c)
${\cal O}_6$, d) ${\cal O}_8$ in units of GeV$^{2d}$, $2d$ being the dimensions of the condensates. The two curves delimit the region
induced by the errors of the data. They co\"\i ncide in almost all regions except the ones above 2.4 GeV$^2$ where the data are
inaccurate.}
\label{fig: malt}
\end{center}
\end{figure*}
%\vspace{-1cm}
\nin
%%%%%%%%%%%%%%%%%%%%%%%%%%%%%%%%%%%%%%%%%%
%%%%%%%%%%%%%%%%%%%%%%%%%%%%%%%%%%%%%%%%%%%%
\section*{d) Finite Energy Sum Rules (FESR)}\label{sec: fesr}
\nin
In \cite{PRADES,DOMINGUEZ,ROJO}, FESR:
\beq\label{eq: fesr}
{\cal M}_n\equiv\int_0^{t_c} dt~t^n \frac{1}{\pi}{\rm Im}\Pi_{LR}=(-1)^n{\cal O}_{2n+2}~, 
\eeq
($n$=0,1,2,3), and its slight variants have been used for determining ${\cal O}_{6,8}$. 
However, unlike the case of LSR analyzed in previous sections, the results depend crucially on the choice of $t_c$ at which one extracts
the optimal results. The two sets of $t_c$-values corresponding to the zeros of
the 1st or/and 2nd Weinberg sum rules are given in Eq.~(\ref{weinberg}). The results from LSR are consistent with the ones
corresponding to value of
$t_c \approx 1.5$ GeV$^2$, while instead in
\cite{PRADES,DOMINGUEZ,ROJO}, the higher value of
$t_c\approx$ 2.5 GeV$^2$ has been favoured. As a consequence, the value of $ O_{8}$ and other higher dimension
condensates obtained in these works are opposite in signs\footnote{We have corrected the sign of 
$O_{8}$ in the curve a), Fig. 5 of \cite{DOMINGUEZ}. Therefore curve a) and b) cross at $t_c\approx 1.3$ GeV$^2$ giving the value of
$O_{8}$ in Table \ref{tab: comparison}.
We have also rescaled the normalization by a factor 2 for consistency in our comparison.} with the ones from LSR and from MHA in large
$N_c$. Taking the value of
$t_c$ in Eq. (\ref{tcoptimal}), we give the version of the FESR results of
\cite{PRADES,ROJO} in  Table
\ref{tab: comparison}, where the slight difference is due to the different parametrizations of the $\tau$-decay data (neural network
in \cite{ROJO}) and to the different weights introduced for improving the original FESR.  
%%%%%%%%%%%%%%%%%%%%%%%%%%%%%%%%%%%%%%%%%%%%%%
\section*{e) Weighted Finite Energy Sum Rules}
\nin
This FESR-like sum rule called ``pinched-weight FESR" (hereafter denoted wFESR) by the authors \cite{MALT}  is an involved
variant of the FESR in Eq. (\ref{eq: fesr}):
\beq\label{eq: pfesr}
J_{\omega_n}\equiv\int_0^{t_c} dt~\omega_n\ga{t\over t_c}\dr \frac{1}{\pi}{\rm Im}\Pi_{LR}~, 
\eeq
where the weight factor $\omega_n$ is:
\beq
\omega_n(x)=x\Bigg{[}1-\ga {n\over n-1}\dr x+\ga{1\over n-1}\dr x^n\Bigg{]}~,
\eeq
for $n=2,3,4,5,6$, and corresponds to the so called maximally safe analysis. The QCD expressions of these sum rules are given in
Eq. (24) of Ref.
\cite{MALT} which we have checked the LO terms. \\
$\bl$ In order to test the results, we study the $t_c$-dependence of
$J_{\omega_1}$ and $J_{\omega_2}$ as shown in Figs. \ref{fig: malt}a) and \ref{fig: malt}b). We include into the analysis, the
known effect of
${\cal O}_4$, which we have also recovered in the previous section. We obtain a
$t_c$-stability point (a compromise region between the convergence of the OPE (small $t_c$) and minimal dependence on the form of
the QCD continuum (large $t_c$)) around 2.0 GeV$^2$, at which we can extract the optimal value of the condensates. However, one can 
notice from Figs. \ref{fig: malt} that contrary to FESR, the estimates are not very sensitive (change in the last digit) to the values of
$t_c$ corresponding to the range in Eq.~(\ref{weinberg}). Neglecting the small
radiative corrections for illustration, one obtains 
 in units of $10^{-3}$ GeV$^6$:
\bea\label{eq: o68} 
J_{\omega_1}&\Lrar&{\cal O}^{(1)}_{68}\equiv {\cal O}_{6}+{3\over 7}{{\cal O}_{8}\over t_c}\approx - 6.6~,\nnb\\
J_{\omega_2}&\Lrar& {\cal O}^{(2)}_{68}\equiv  {\cal O}_{6}+{1\over 2}{{\cal O}_{8}\over t_c}\approx -5.8~ .
\eea
We insert into this expression the values in units of $10^{-3}$ of ${\cal O}_{6,8}$ fitted by \cite{MALT} (mean value from ALEPH and
OPAL fit):
\beq\label{eq: o6+8}
{\cal O}_{6}\simeq -4.9~,~~~~~{\cal O}_{8}\simeq -3.8~,
\eeq
which gives: 
\beq 
{\cal O}^{(1)}_{68}(2.15)\simeq -5.7 ~~~~{\rm and}~~~{\cal O}^{(2)}_{68}(2.15)\simeq -5.8~. 
\eeq 
This test shows the
consistency between the results obtained using
$t_c$ stability in Eq. (\ref{eq: o68}) and the least square fit in Eq. (\ref{eq: o6+8}).\\
$\bl$  Alternatively, we can also
solve the two equations
$J_{\omega_1}$ and
$J_{\omega_2}$ for extracting the two solutions ${\cal O}_{6}$ and ${\cal O}_{8}$. We study the $t_c$-dependence of the
results in Figs. \ref{fig: malt}c) and \ref{fig: malt}d). Here the stability is obtained at $t_c\simeq 1.7$ GeV$^2$ which
differs from the one obtained previously. We may interpret this difference as due to the fact that we do not 
consider here the same observables as in Figs.
\ref{fig: malt}a) and 
\ref{fig: malt}b). However, in order to have conservative results, we shall consider $t_c$ in the range ($1.7\sim 2.0$) GeV$^2$
where the two stabilities are obtained. One can inspect that, in this range, the estimate is only slightly affected by the $t_c$-values. 
The results are given in Table~\ref{tab: comparison} with a good accuracy. We check again the consistency of the results by 
inserting these values
into Eq. (\ref{eq: o68}), which gives:
\beq
{\cal O}^{(1)}_{68}(2.15)\simeq -6.7~~~{\rm and}~~~{\cal O}^{(2)}_{68}\simeq -6.1~. 
\eeq
$\bl$ Our test does not support the results given in
\cite{MALT} obtained from numerical fits. This may due to the fact that the terms in the series have alternate signs, and/or where
the 2nd term is a small correction of the 1st one, and may be difficult to extract from the fitting procedure . Instead, we expect that
the new results from this method which we give in Table
\ref{tab: comparison}  obtained using stability criteria, from solving the two equations $J_{\omega_1}$ and
$J_{\omega_2}$ for extracting the two unknown ${\cal O}_{6}$ and ${\cal O}_{8}$ are more 
reliable.\\
$\bl$ Notice that in a large range of $t_c$, the two estimates of ${\cal O}_{6}$ and ${\cal O}_{8}$ do not flip sign, which,
like in the case of LSR, can be due to the weight  factor in the spectral integral. This is not the case of the basic FESR. \\ 
$\bl$ In
principle, once one knows
${\cal O}_{6}$ and
${\cal O}_{8}$, one can extract the other high-dimension condensates from the set of equations given in Eq. (24) of
\cite{MALT}. ${\cal O}_{10}$ to ${\cal O}_{16}$ can be, e.g., extracted from $J_{\omega_3}$ to
$J_{\omega_{10}}$. However, more we go to higher moments, less the accuracy on the estimate is reached as the
high-dimension terms which one wishes to extract are tiny corrections to the leading order terms, while the method is
not accurate enough to pick up these tiny corrections. For
instance at $t_c\approx 2$ GeV$^2$, the QCD parts of the sum rules normalized to the leading ${\cal O}_6$ contributions read:
\bea\label{eq: j14}
J_{\omega_5}&\sim& \# \Big{[} 1+0.005\ga{2~\rm{GeV}^2\over t_c}\dr^4 {\cal O}_{14}\Big{]}~,\nnb\\
J_{\omega_6}&\sim& \# \Big{[} 1-0.002\ga{2~\rm{GeV}^2\over t_c}\dr^5  {\cal O}_{16}\Big{]}~.
\eea
The corrections are a factor 2 smaller for $J_{\omega_9}$ and $J_{\omega_{10}}$. This fact may explain why relatively large
central values of the high-dimension condensates emerge from this method. A tentative  extraction of 
${\cal O}_{10}$ to ${\cal O}_{16}$ from  $J_{\omega_3}$ to
$J_{\omega_6}$ shows that the $t_c$-dependence present a flat stability around 1.25 GeV$^2$ and another extremum around
1.8 GeV$^2$. The values obtained at the second point are very sensitive to the input value of ${\cal O}_{6}$ and flip
sign compared to the one at the flat plateau for $D\geq 14$, a feature similar to ${\cal O}_8$ from
FESR analysis \cite{PRADES,ROJO}. This may indicate that the weight factor is less efficient for high-dimension condensates.
We have excluded the
high-$t_c$ solution similarly to the FESR case, and we deduce the values in Table~\ref{tab: comparison}.
%%%%%%%%%%%%%%%%%%%%%%%%%%%%%%%%%%%%%%%%%%%%%%%%%%
\subsection*{f) Test of the factorization assumption}
\nin  
$\bl$ The $D=6$ condensate contributions to $\Pi^{LR}$ have been first derived in \cite{SNWEIN} using the leading
order result of
\cite{SVZ}  for the vector and axial-vector correlators. The radiative corrections have been obtained in
\cite{CHET2,CHET3}. Using an anti-commuting $\gamma_5$ matrix and the choice of operator basis in \cite{CHET3}, it reads
by assuming a factorisation of the four-quark condensates :
\beq
{\cal O}_6=-{64\over 9}\pi\alpha_s\la\bar uu\ra^2\Bigg{[} 1+{\alpha_s\over \pi}\ga {89\over 48}-{1\over 4}\ln
{Q^2\over\nu^2}\dr\Bigg{]}.
\eeq
Using the  NDLR or/and the HV regularization scheme, the same contribution reads, to leading order in $N_c$
at $Q^2=\nu^2$ \cite{PRADES}:
\bea
{\cal O}_6=-{8}\pi\alpha_s\Big{[}\la\bar uu\ra^2\ga 1+{\alpha_s\over \pi}{61\over 12}\dr
-
\frac{1}{16\pi^2}A_{LR}\Big{]},
\eea
where $A_{LR}\simeq (4.4\pm 0.5)\times 10^{-3}$ is of order $\alpha_s^2$. \\
$\bl$ The $D=8$
four-quark condensate contributions have been obtained in
\cite{SMILGA} where a
$1/N_c^2$ ambiguity has been noticed. The $D=10$ condensates have been obtained in \cite{ZYA}.
Assuming factorization, one can write:
\bea
{\cal O}_8&=&{64\over 9}\pi\alpha_s\la\bar uu\ra^2M_0^2~,\nnb\\
{\cal O}_{10}&=&-{8\over 9}\pi\alpha_s\la\bar uu\ra^2\Big{[}{50\over 9}M_0^2+32\pi\la\alpha_s G^2\ra\Big{]}~.
\eea
$\bl$ $M_0^2$ is the scale gouverning
the mixed condensate and is equal to $(0.8\pm 0.2)$ GeV$^2$ from the
baryon sum rules
\cite{DOSCH,JAMIN}, $B$-$B^*$ mass-splitting \cite{SNHEAVY} and string model \cite{DIGIACOMO}. We shall use the  value of the
gluon condensate $\la \alpha_s G^2\ra=(0.07\pm 0.01)$ GeV$^4$ from $e^+e^-$ data
\cite{SNG,SNH}. Within the factorization assumption, we shall include the $\log$-dependence of the quark
condensate and of $\alpha_s$, which give:
\bea
\alpha_s \la\bar uu\ra^2|_{fac}\simeq\frac{2}{9}\frac{\pi}{(\log{Q/\Lambda})^{1/9}}\la\widehat{\bar uu}\ra^2~,
\eea
which is $1.6\times
10^{-4}$ GeV$^6$ if one uses the invariant quark condensate $\la\widehat{\bar uu}\ra\simeq -(248~{\rm MeV})^3$
\cite{SNMASS} and evaluate $\alpha_s(\tau)$ at $\tau=1.5$ GeV$^{-2}$ at which ${\cal O}_8$ has been extracted from
the sum rule.  Using these numerical inputs, we deduce in units of $10^{-3}$GeV$^{D}$ ($D$ being the dimension of the
condensate):
\bea
{\cal O}_6|_{fac}&\approx&-3.6~,\nnb\\
{\cal O}_8|_{fac}&=& +2.9~,\nnb\\
{\cal O}_{10}|_{fac}&=&-4.7~.
\eea
$\bl$ Comparing these values with the ones in Table 1 and section \ref{sec: improve}, we conclude that the
factorization assumption agrees in sign with these results but underestimate the absolute value of the condensates
by a factor 2-5. This feature is similar to the case of the vector \cite{TARRACH,BERTLMANN,SNG}, axial-vector
\cite{PAPA,ZAKA2}, baryon
\cite{JAMIN}  sum rules and from the analysis of the $V$ and $V+A$ $\tau$-decay data \cite{ALEPH,OPAL,ZAKA2}. From the theoretical
point of view, the factorization assumption is only consistent with the renormalization of operators to leading order in $1/N_c$
due to mixing of different operators having the same dimensions \cite{TARRA}.
%\vspace{-.5cm}
%%%%%%%%%%%%%%%%%%%%%
\section{CONCLUSIONS}
\nin
$\bl$ We have used the $V$-$A$ component of the hadronic tau decays data for exploring the vacuum
structure of the $\Pi^{LR}$ QCD correlator using a set of Laplace sum rules (LSR). We have also revisited different estimates
based on FESR and its variant in section \ref{sec: comparison}. Our results are summarized in Table \ref{tab: comparison}. 
\\
$\bl$ Contrary to most papers in the literature, we do not perform a least square fitting procedure for extracting simultaneously
different condensates, but instead use the stability criteria (existence of minima or inflexion points) for our estimate of the condensates.
Due to the alternate signs of the condensate contributions in the OPE and to the fact that in most methods, the high-dimension condensate
contributions are corrections to the lowest dimension condensates in the analysis, the approaches for extracting these high-dimension
condensates can become inaccurate.
\\
$\bl$ Instead, our strategy is to look for sum rules which disantangle, from the beginning, the relevant
high-dimension condensates, and then makes the analysis cleaner and more transparent.\\
$\bl$ We have given a first estimate of the size of the $D=18$ condensates, which will be interesting to check using
alternative methods. The LSR estimate, which we expect to be more appropriate for extracting higher dimension condensates 
than wFESR and FESR, shows that
the size of the very high
$D=16$ and
$D=18$ condensates are relatively small which may indicate the good convergence of the OPE even at large $\tau$-values.\\
$\bl$ During the analysis, as one can see in previous figures, the absolute values of the condensates are slightly affected by
$\tau$ and
$t_c$ in the optimum region (minimum or inflexion point). However, except for $D=4$, it is important to notice
that the results from LSR in large range of $\tau$- and $t_c$-values do not flip
sign, which is a great advantage compared to the ones from some Finite Energy like-Sum Rules
discussed in the literature.
\\
$\bl$ The extension of the present analysis to some other channels are feasible though not straightforward. This is due to the
relative importance of the continuum pQCD contribution for higher moments in some other channels, which is not the case of $V$-$A$,
where this effect exactly cancels at higher energies. We plan to come back to these different channels in a future
publication.
%%%%%%%%%%%%%%%%%%%%%%%%%
\section*{ACKNOWLEDGEMENT}
\nin
We thank Eduardo de Rafael for useful comments. 
%%%%%%%%%%%%%%%%%%%%%%%%%%%

\end{document}